\documentclass[12pt,preprint]{aastex}
\def\degpoint{\ifmmode ^{\rm{o}}\!. \else $^{\rm{o}}\!.$\fi}

\newcommand{\kms}{\mbox{km\,s$^{-1}$}}
\newcommand{\Msun}{\mbox{$M_{\odot}$}}
\newcommand{\Rsun}{\mbox{$R_{\odot}$}}

\newcommand{\Lsun}{\mbox{$L_{\odot}$}}

\begin{document}

\title{The Pan-Pacific Planet Search V. Fundamental Parameters for 164 
Evolved Stars }

\author{Robert A.~Wittenmyer\altaffilmark{1,2,3}, Fan Liu\altaffilmark{4}, 
Liang Wang\altaffilmark{5,6}, Luca Casagrande\altaffilmark{4}, John Asher 
Johnson\altaffilmark{7}, C.G.~Tinney\altaffilmark{1,2} }

\altaffiltext{1}{School of Physics, UNSW Australia, Sydney 2052, 
Australia}
\altaffiltext{2}{Australian Centre for Astrobiology, UNSW Australia, 
Sydney 2052, Australia}
\altaffiltext{3}{Computational Engineering and Science Research Centre, 
University of Southern Queensland, Toowoomba, Queensland 4350, 
Australia}
\altaffiltext{4}{Research School of Astronomy \& Astrophysics, 
Australian National University, Cotter Road, Weston Creek, ACT 2611 
Australia}
\altaffiltext{5}{Key Laboratory of Optical Astronomy, National 
Astronomical Observatories, Chinese Academy of Sciences, A20 Datun Road, 
Chaoyang District, Beijing 100012, China}
\altaffiltext{6}{Max-Planck-Institut f\"ur Extraterrestrische Physik, 
Giessenbachstrasse, 85748 Garching, Germany}
\altaffiltext{7}{Harvard-Smithsonian Center for Astrophysics, Cambridge, 
MA 02138 USA}
\email{rob@unsw.edu.au}

\shorttitle{PPPS Stellar Parameters }
\shortauthors{Wittenmyer et al.}


\begin{abstract}

\noindent We present spectroscopic stellar parameters for the complete 
target list of 164 evolved stars from the Pan-Pacific Planet Search, a 
five-year radial velocity campaign using the 3.9m Anglo-Australian 
Telescope.  For 87 of these bright giants, our work represents the first 
determination of their fundamental parameters.  Our results carry 
typical uncertainties of 100\,K, 0.15 dex, and 0.1 dex in $T_{\rm eff}$, 
$\log g$, and [Fe/H] and are consistent with literature values where 
available.  The derived stellar masses have a mean of 
$1.31^{+0.28}_{-0.25}$\,\Msun, with a tail extending to $\sim$2\,\Msun, 
consistent with the interpretation of these targets as ``retired'' A-F 
type stars.

\end{abstract}



\section{Introduction}

Understanding the target stars is critical to any planet search.  
Knowing the stellar physical parameters (most critically, the mass and 
radius) is of course necessary for further characterisation of any 
planets found, but this information is also important for placing the 
complete results -- detections and non-detections -- into context.  For 
example, one result arising from studies of evolved stars is a relative 
deficit of short-period planets, despite obvious selection biases in 
favor of detecting them.  This apparent shortfall has been noted by 
\citet{johnson07} and \citet{sato10}.  Two possible explanations are 
that either the planets are absent, or they are swallowed by the host 
star as it expands \citep{kunitomo11}.  We are currently testing the 
first hypothesis by making high-cadence observations of selected giants 
using the Weihai Observatory 1m telescope \citep{weihai, gao14, guo14, 
hu14}.  Testing the second hypothesis requires accurate measurements of 
the stellar radii.  As most of these evolved stars are usually too 
distant for direct measurement via interferometry 
\citep[e.g.][]{ligi12,b13}, we must rely on spectroscopic determinations 
of effective temperatures, and model-derived luminosities to arrive at 
the radii.  We note that some brighter giants have had asteroseismic 
radius determinations based on \textit{Kepler}/K2 photometry 
\citep{stello15}, and future spacecraft missions such as TESS and PLATO 
will provide additional direct measurements.

A positive correlation between giant planet occurrence and host-star 
metallicity has been well-established for main-sequence stars 
\citep{gonzalez97, fv05}.  \citet{johnson10a} found the 
planet-metallicity correlation to hold for subgiants from their Lick and 
Keck survey.  However, the situation for giant stars is far less clear.  
No correlation was found for G and K giants by \citet{pasquini07}, 
\citet{takeda08}, and \citet{mortier13}, whereas \citet{hekker07} found 
a positive correlation in their sample of K giants.  \citet{maldonado13} 
obtained mixed results, with a planet-metallicity correlation only 
evident for subgiants and giants with $M_* > 1.5$\,\Msun.  A recent 
study of 12 years of precise radial velocity data on 373 G/K giants by 
\citet{reffert15} revealed a strong correlation.  An analysis of a 
subsample with uniform planet detectability gave the same result, giving 
confidence that the observed planet-metallicity correlation is not a 
product of biases in the sample.

The Pan-Pacific Planet Search (PPPS) operated at the 3.9\,m 
Anglo-Australian Telescope (AAT) from 2009-2014, targeting 164 Southern 
Hemisphere evolved stars \citep{47205paper}.  The PPPS targets are 
redder than those observed by most surveys \citep{mortier13} -- we have 
chosen stars with $1.0\le(B-V)\le\,1.2$), whereas other surveys enforce 
$(B-V)\le\,1.0$.  This colour selection makes the PPPS targets 
complementary to the $\sim$450 Northern ``retired A stars'' from the 
well-established Lick and Keck program \citep{johnson06, johnson11}.  A 
complete target list is given in \citet{47205paper}.  In this work, we 
present fundamental parameters for all PPPS targets as derived from 
high-resolution, high signal-to-noise spectra obtained in the course of 
the planet search program.


\section{Observations}

All observations were carried out at the AAT using its UCLES echelle 
spectrograph \citep{diego:90}.  The PPPS program uses the Doppler 
technique for measuring precise radial velocities, with an iodine 
absorption cell to calibrate the spectrograph point-spread-function 
\citep{val:95,BuMaWi96}.  An iodine-free ``template'' observation is 
acquired for each target at a resolution $R\sim$60,000 and a 
signal-to-noise of 100-300 per pixel.  The radial velocity of each star 
is then measured relative to the zero-point defined by its template 
\citep{47205paper, 121056paper, 155233}.  In this work, we use the 
iodine-free templates to determine spectroscopic stellar atmospheric 
parameters.


\section{Stellar Parameter Determination}

\subsection{Spectroscopic Method}

\begin{figure}
\centering
\includegraphics[width=0.85\textwidth]{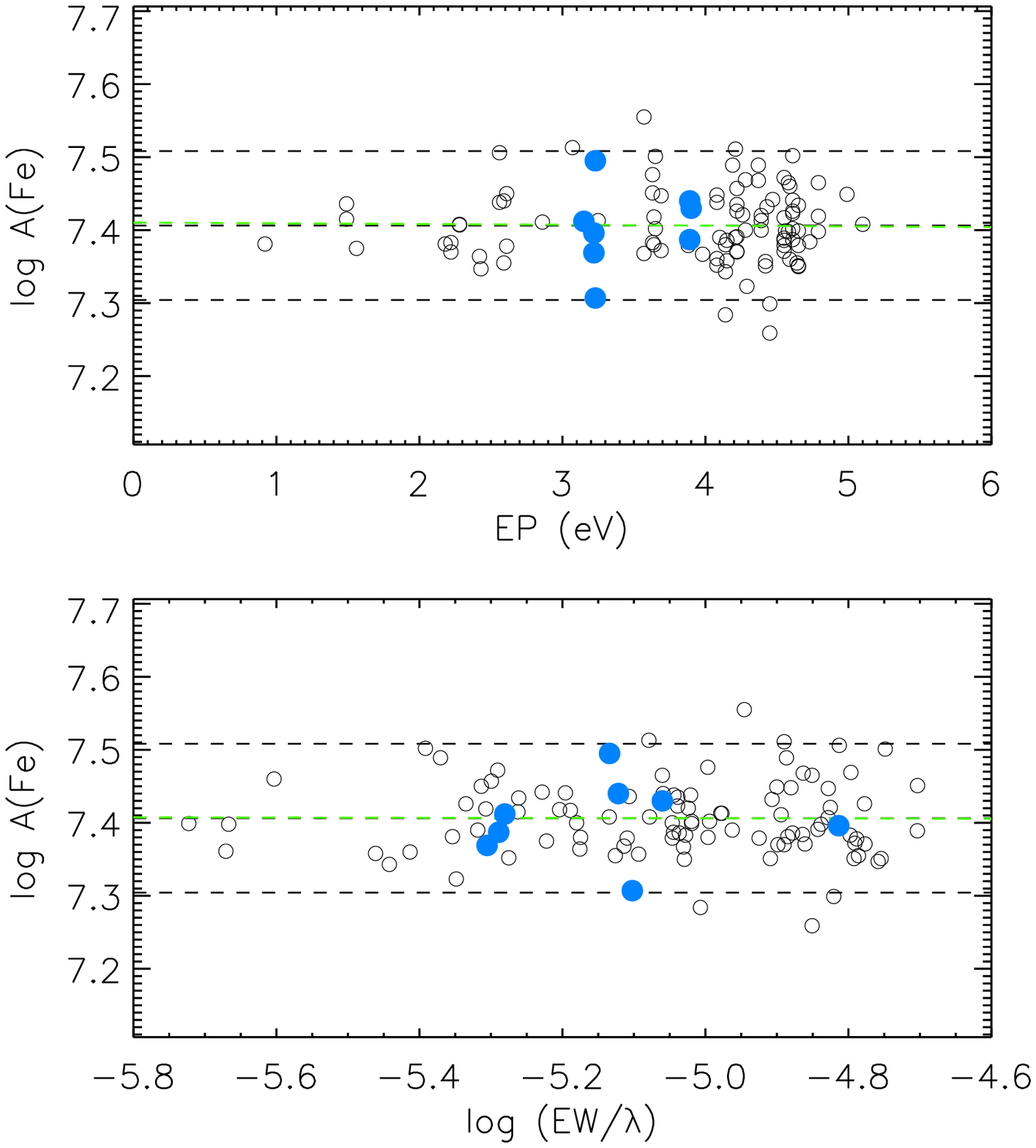}
\caption{Top panel: log\,A(Fe) of a typical sample star (HD 206993) 
derived as a function of excitation potential; open circles and blue 
filled circles represent Fe\,{\sc i} and Fe\,{\sc ii} lines, 
respectively.  The green dashed line shows the location of the mean 
log\,A(Fe), while black dashed lines represent twice the standard 
deviation, $\pm 2 \sigma$.  Bottom panel: same as in the top panel but 
as a function of reduced equivalent width.}
\label{excite}
\end{figure}

\begin{figure}
\centering
\plottwo{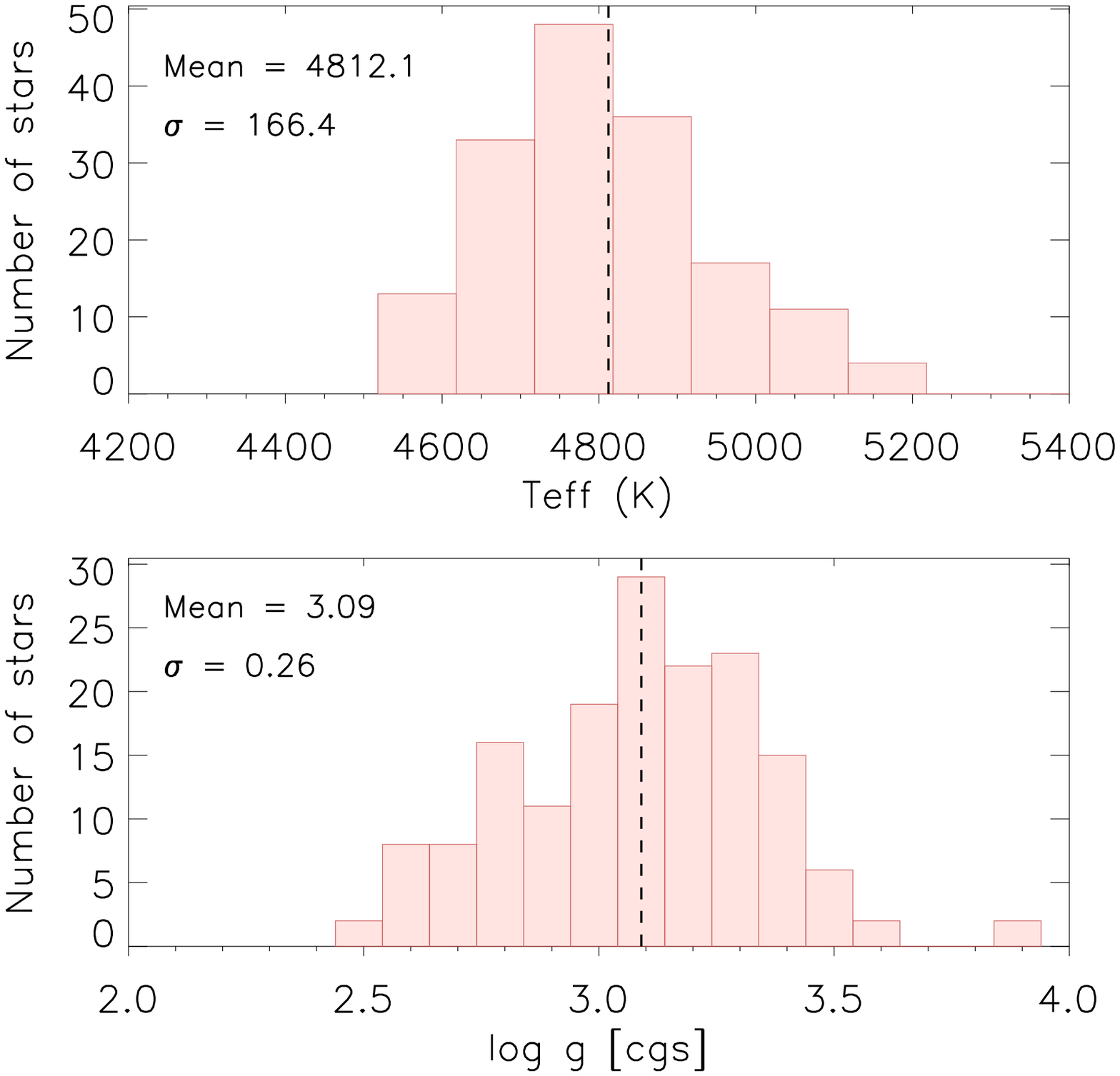}{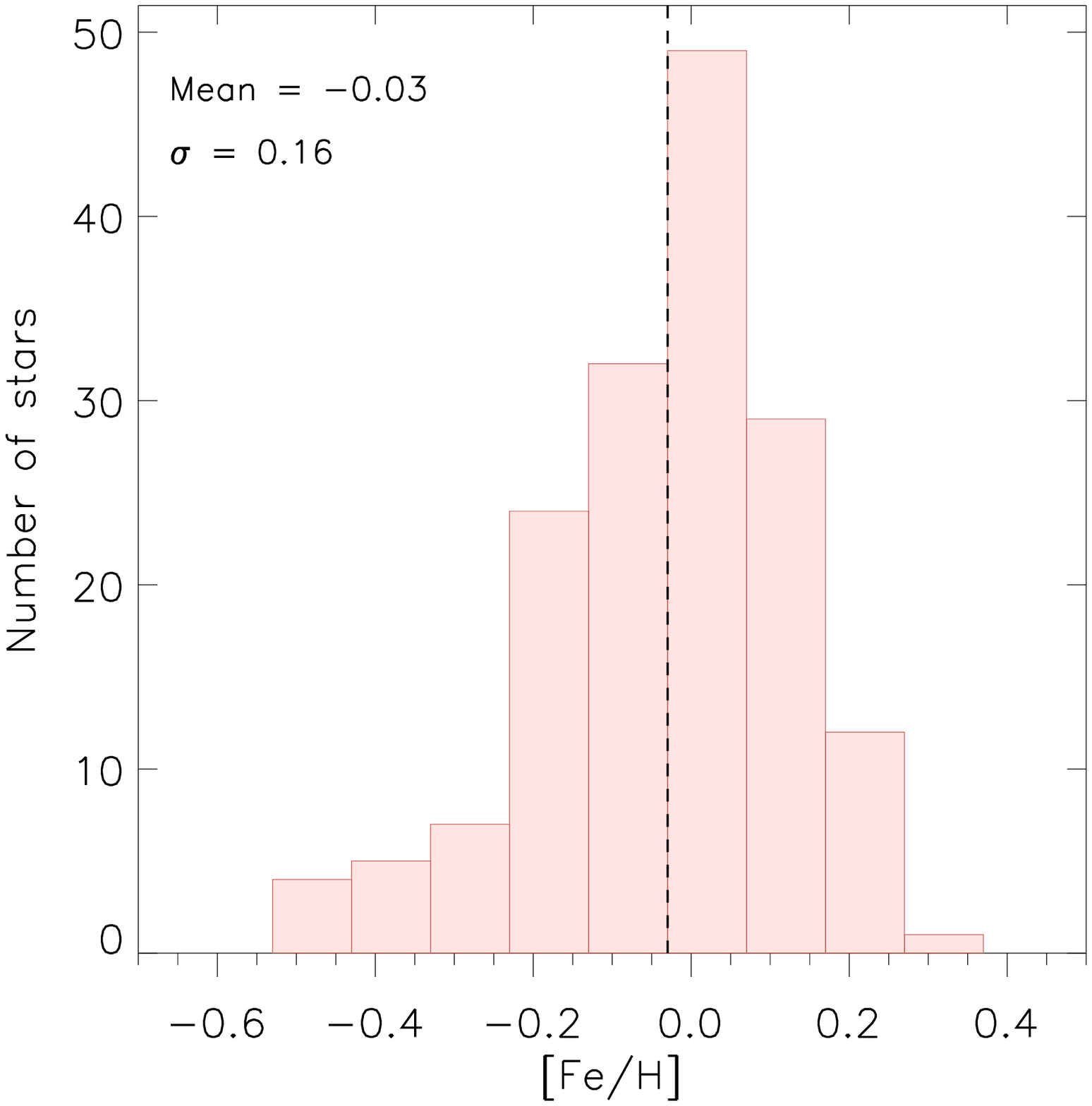}
\caption{Left panel: distributions of spectroscopic T$_{\rm eff}$, $\log g$ 
of our sample stars. Right panel: distributions of [Fe/H] of our sample 
stars.}
\label{distrib}
\end{figure}

We started our analysis by automatically measuring the equivalent widths 
(EWs) of the spectral lines using the ARES code 
\citep{sou07}\footnote{The parameter 'rejt' in the code was set to be 
REJT = 1.0 $-$ 1.0/(S/N), which is 0.992 for our sample.}.  The line 
list employed in our analysis was adopted from \citet{ts13}.  Lines too 
weak ($<$ 5 m\AA) or strong ($>$ 110 m\AA) were excluded from the 
analysis. Then we addressed a standard 1D, local thermodynamic 
equilibrium (LTE) abundance analysis using the 2013 version of MOOG 
\citep{sne73} with the ODFNEW grid of Kurucz ATLAS9 model atmospheres 
\citep{cas03}.  In order to determine the stellar parameters (effective 
temperature $T_{\rm eff}$, surface gravity $\log g$, microturbulence 
$\xi_t$ and metallicity [Fe/H]), we force the excitation/ionization 
balance by minimizing the slopes in log\,A(Fe I) versus lower excitation 
potential (EP) and reduced EW ($\log (EW/\lambda$)) as well as the 
difference between log\,A(Fe I) and log\,A(Fe II), simultaneously.  We also 
require the derived average metallicity to be consistent with the adopted 
model atmospheric value.  We adopted the final results by iterating the 
whole process until the balance is exactly achieved.  Lines whose 
abundances departed from the average by $>$ 3$\sigma$ were clipped 
during the analysis. We adopted the solar values from \citet{asp09} as a 
zero point.  The stellar spectroscopic parameters of our sample stars 
are listed in Table~\ref{spec}.  Figure~\ref{excite} shows the resulting 
excitation and ionization balance of a typical sample star (HD\,206993).  
By adding perturbations of each parameter to change the slopes or 
abundance difference within a reasonable range, we are able to 
conservatively estimate the typical uncertainties of $T_{\rm eff}$, 
$\log g$, $\xi_t$ and [Fe/H] of our sample stars to be $\sim$ 100 K, 
0.15 dex, 0.15 \kms\ and 0.1 dex, respectively.  Since this sample has 
been chosen to lie in a specific region of the H-R diagram such that 
they are all in a similar evolutionary state, we expect there to be 
little variation in uncertainties from star to star.  Hence we have 
given conservative uncertainty estimates for the whole sample.  The mean 
spectroscopic $T_{\rm eff}$ of the sample is 4812 K with a standard 
deviation ($\sigma$) of 166 K, while $\langle\log g\rangle$ = 3.09 $\pm$ 
0.26. The average [Fe/H] of the sample is $-$0.03 $\pm$ 0.16, which is 
slightly more metal-poor than the solar metallicity.  We plot the 
distributions of spectroscopic parameters of our sample stars in 
Figure~\ref{distrib}.

\subsection{Photometric Method}

\begin{figure}
\centering
\plottwo{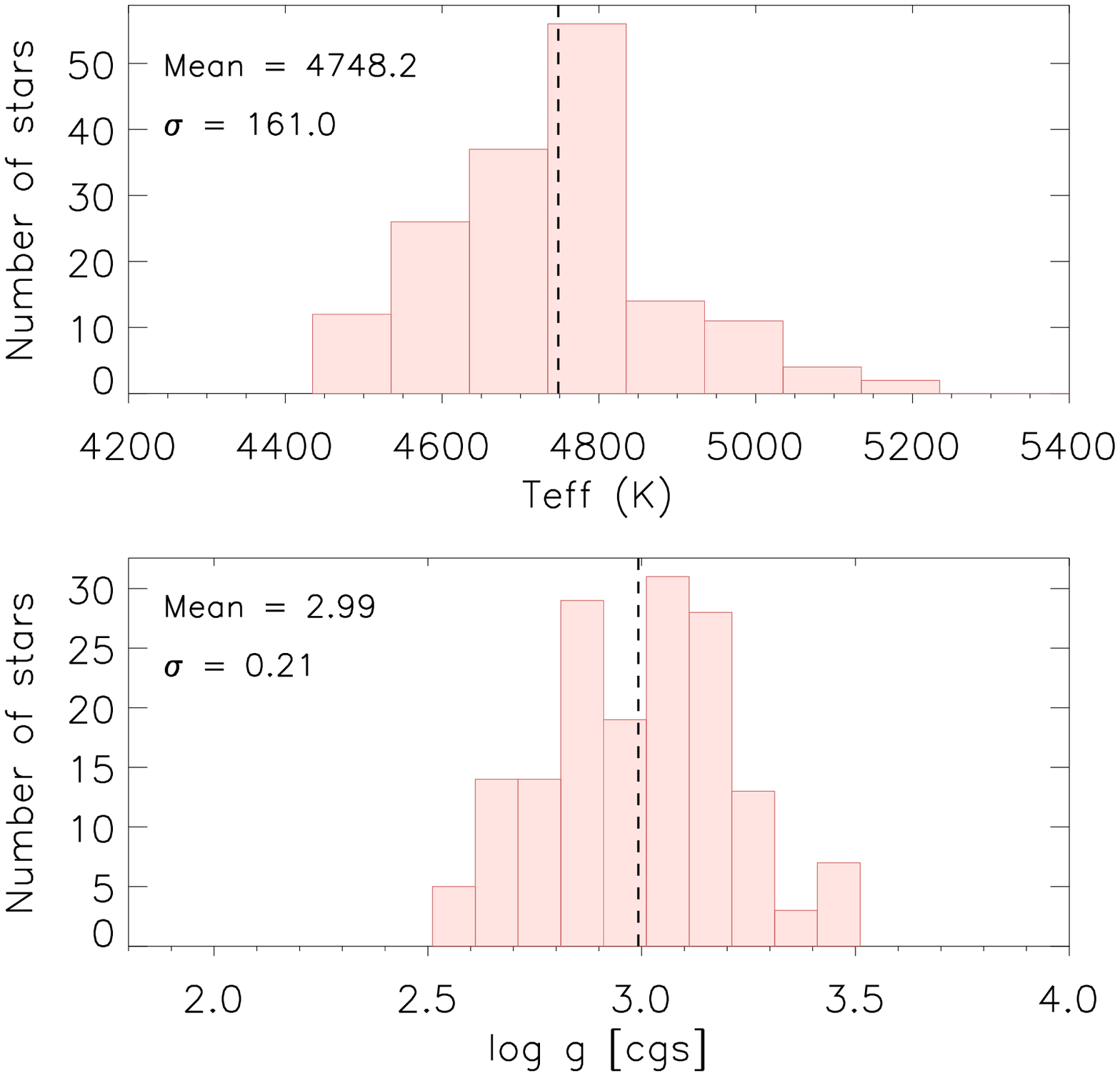}{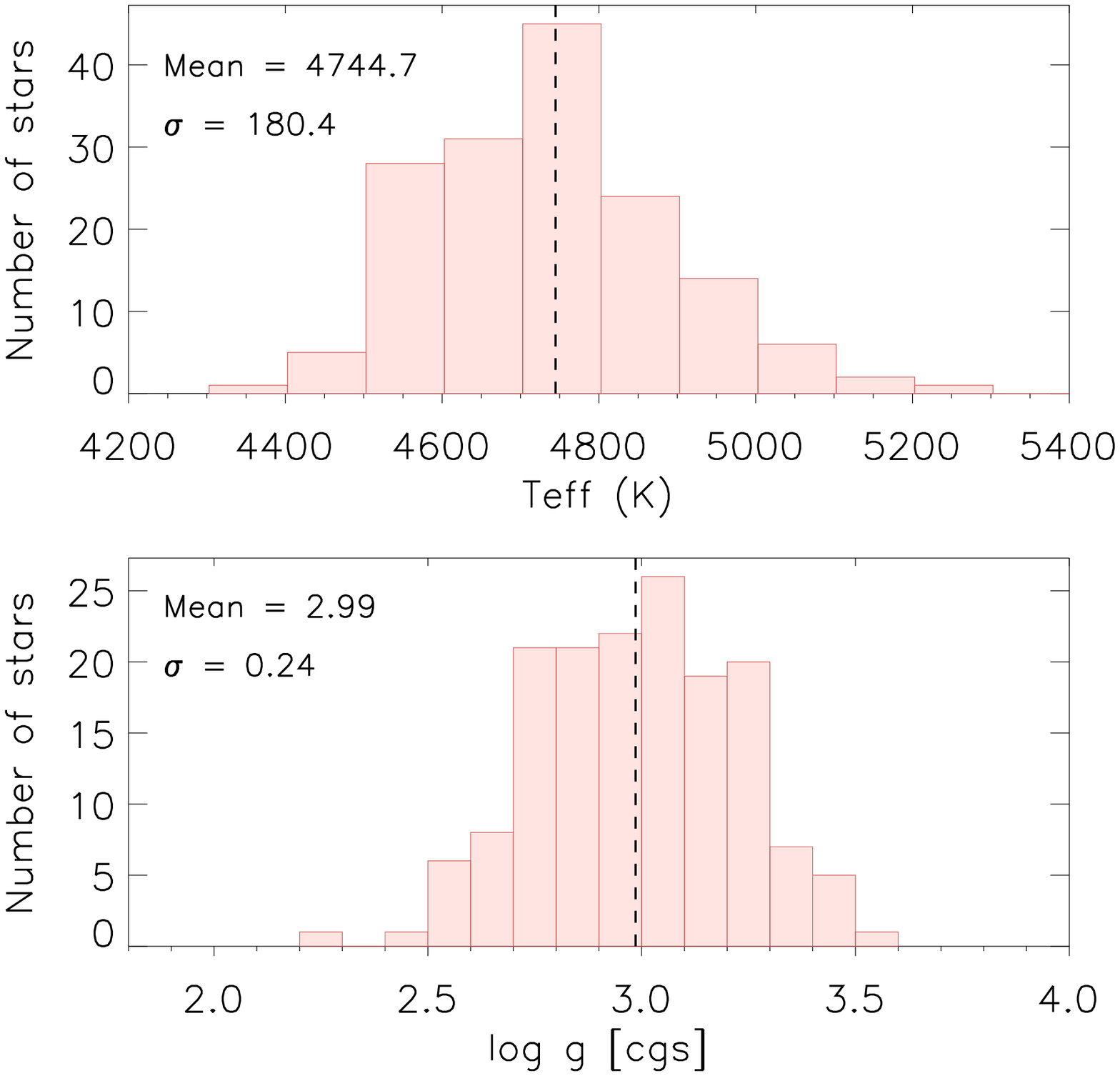}
\caption{Left panel: distributions of T$_{\rm eff}$(B - V), $\log g$(B - V) 
of our sample stars. Right panel: distributions of T$_{\rm eff}$(V - K), 
$\log g$(V - K) of our sample stars.}
\label{fig3}
\end{figure}

We derived the effective temperature ($T_{\rm eff}$) of our sample stars 
from the $(B - V)$ and $(V - K)$ photometric data, using the empirical 
calibration relations from \citet{am99}\footnote{The choice of 
relationships depends on which region $(B - V)$ or $(V - K)$ falls on 
for individual programme star.}.  These photometric parameters are given 
in Table~\ref{phot}.  We plot the histograms of photometric 
parameters of our sample stars in Figure~\ref{fig3}.  Both methods show 
very similar distributions.  The $B$, $V$ and $K$ colour indices were 
obtained from the SIMBAD database.  We adopted the reddening estimation 
according to \citet{sch98} with the corrections stated by \citet{ag99} 
and \citet{bee02} to obtain the colour excess $E(B - V)_{\rm A}$.  For 
nearby stars, the reddening value is calculated as: $E(B - V) = [1 - 
\exp (-|D\sin b|/125)]E(B - V)_{\rm A}$, where $D$ is the distance of 
the star and $b$ is the Galactic latitude, both were obtained from the 
SIMBAD database.  Then, we adopted $E(V - K) = 2.948 E(B - V)$ as the 
colour excess for $(V - K)$ \citep{sch98}.  The values of reddening are 
listed in Table~\ref{reddening}.

The surface gravity ($\log g$) was estimated with the method described 
by Liu et al. (2007, 2012) with the equations below:
\begin{equation}
\log g = \log g_\odot + \log\left(\frac{M}{M_\odot}\right) + 
4 \log\left(\frac{T_{\rm eff}}{T_{\rm eff,\odot}}\right) + 0.4 (M_{\rm bol} - M_{\rm bol,\odot})
\end{equation}
\begin{equation}
M_{\rm bol} = V + BC + 5 \log \pi + 5 - A_V
\end{equation}
\begin{equation}
A_V = 3.1 E(B - V)
\end{equation}

Here, $T_{\rm eff}$ are the temperatures derived using the photometric 
method, $M_{\rm bol}$ are the bolometric magnitudes, and $V$, $BC$, $\pi$ and 
$A_V$ represent the apparent $V$ magnitude, bolometric correction, 
parallax and interstellar extinction, respectively.  We note that the 
bolometric corrections ($BC$) are calculated based on \citet{am99}, using 
photometric temperatures and metallicities derived with spectroscopic 
method.  The parallaxes $\pi$ are taken from the SIMBAD database.
Stellar masses, ages, radii, and luminosities are estimated by finding 
the best match of derived ($T_{\rm eff}$, $M_{\rm bol}$) to the values 
predicted by theoretical evolutionary models with given [Fe/H] 
\citep[e.g.][]{wang2011}.  We adopt the Yale-Yonsei (Y$^2$) tracks with 
an improved core overshoot treatment \citep{yi03, demarque04}, and use a 
Newtonian polynomial to interpolate between that grid.

\begin{figure}
\centering
\includegraphics[width=0.85\textwidth]{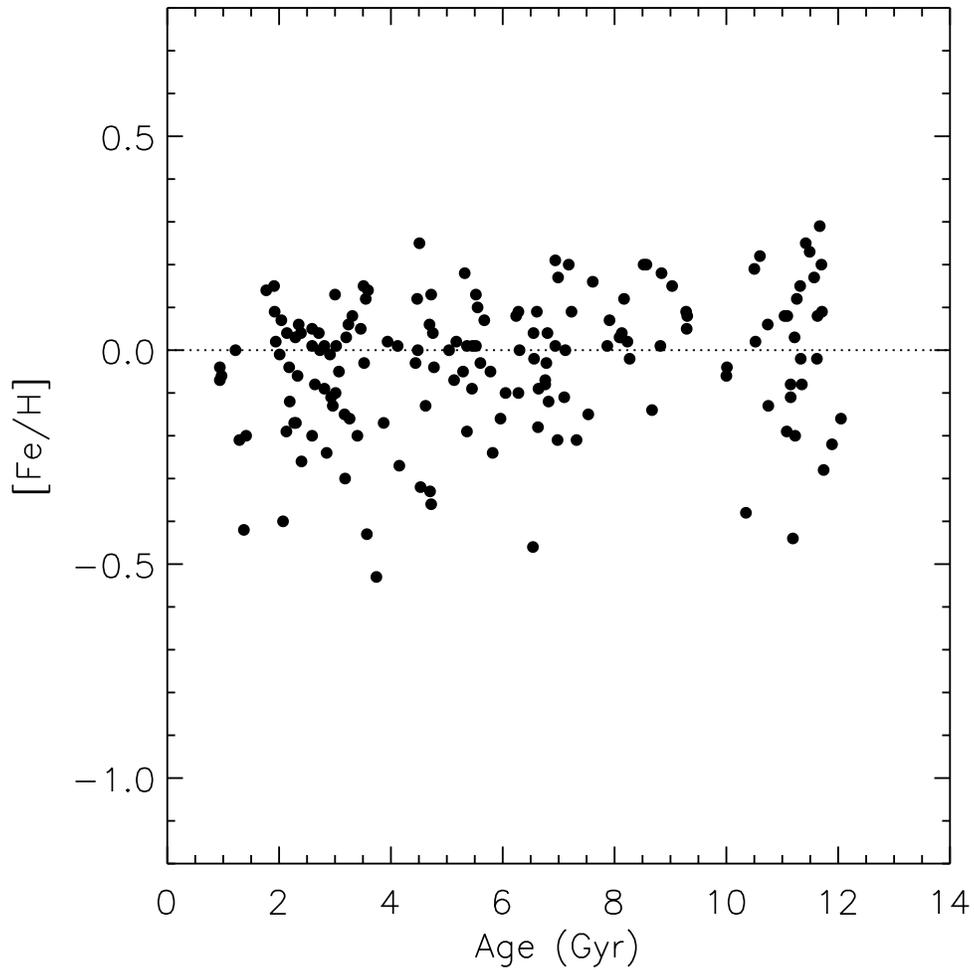}
\caption{Metallicity [Fe/H] versus age for our sample, indicating a flat 
relation consistent with the Solar neighbourhood \citep{haywood08, 
casa11}.  The slope is 0.010$\pm$0.004 with an rms scatter of 0.154 
about the fit. }
\label{age}
\end{figure}

\begin{figure}
\centering
\includegraphics[width=0.85\textwidth]{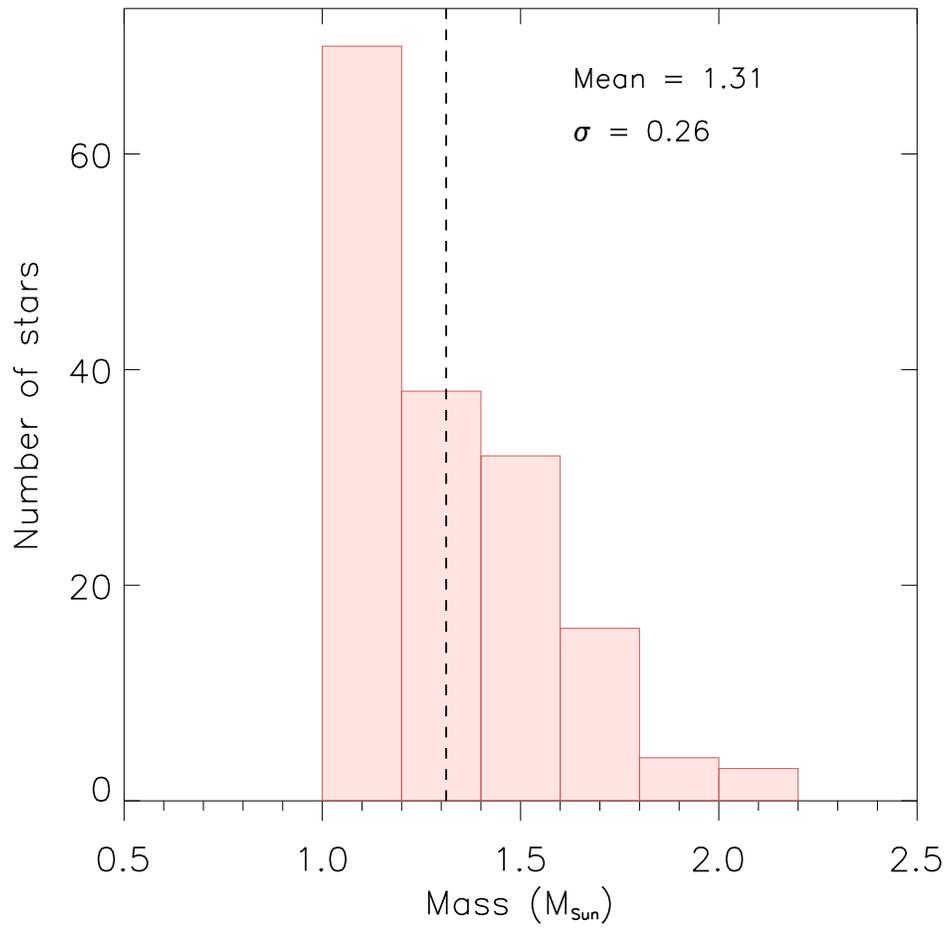}
\caption{Distributions of stellar mass of our sample stars.}
\label{fig4}
\end{figure}

Our derived stellar parameters (mass, luminosity, radius, age) 
are given in Table~\ref{masses}.  Typical uncertainties are 
0.15-0.25\,\Msun\ and 0.5-0.6\,\Rsun.  Figure~\ref{age} shows the 
age-metallicity relation for this sample, indicating a flat distribution 
which is consistent with the Solar neighbourhood.  We also plot the 
distributions of stellar mass of the whole sample in Figure~\ref{fig4}, 
which indicate that our sub-giants sample is well represented with a 
mean mass of $1.31^{+0.28}_{-0.25}$\,\Msun. 

\begin{figure}
\centering
\includegraphics[width=0.85\textwidth]{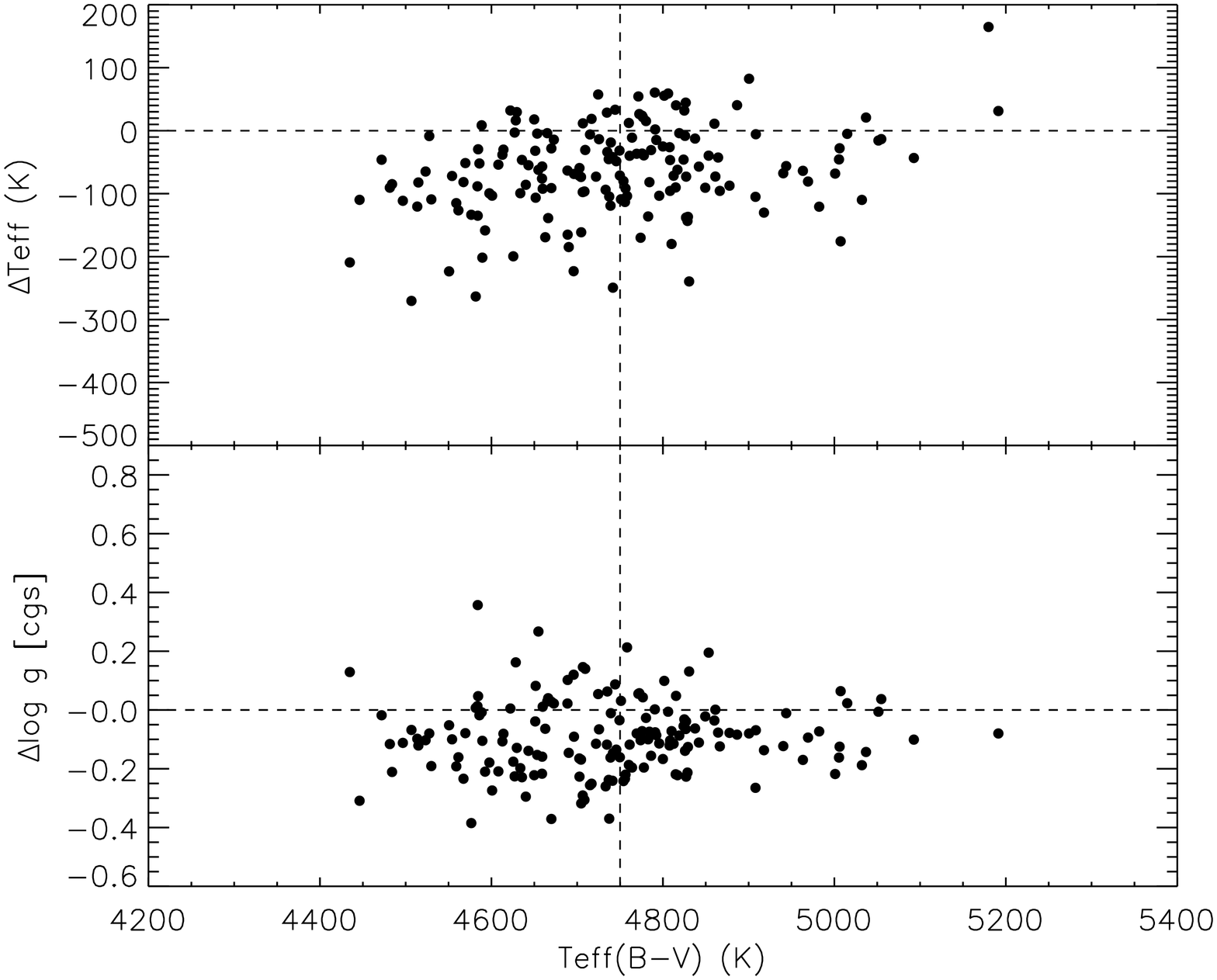}
\caption{Upper panel: T$_{\rm eff}$(B - V) minus T$_{\rm eff}$(spec) as a 
function of T$_{\rm eff}$(B - V). Lower panel: $\log g$(B - V) - $\log 
g$(spec) as a function of T$_{\rm eff}$(B - V).}
\label{fig5}
\end{figure}
\begin{figure}
\centering
\includegraphics[width=0.85\textwidth]{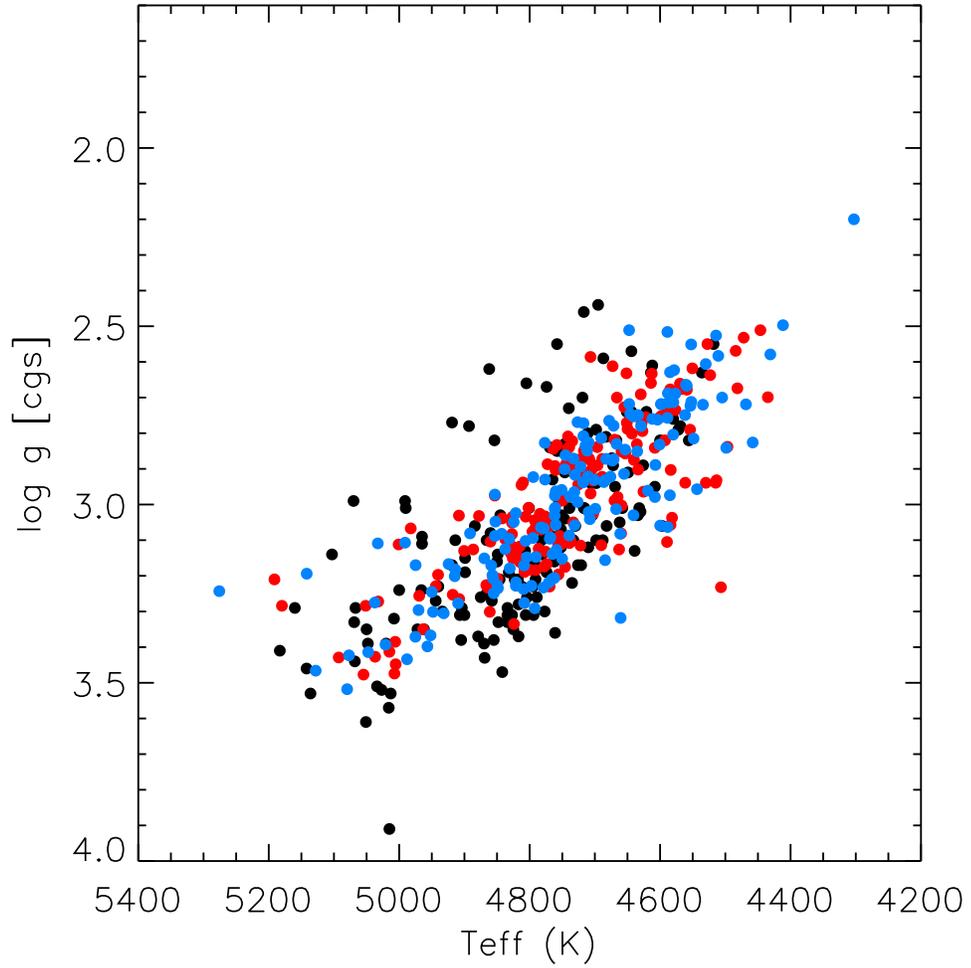}
\caption{$\log g$ versus T$_{\rm eff}$ derived with spectroscopic and 
photometric methods; black, blue and red filed circles represent the 
results derived from the spectroscopic method, photometric method (B - V) 
and photometric method (V - K), respectively.}
\label{fig6}
\end{figure}

Figure~\ref{fig5} compares derived $T_{\rm eff}$ and $\log g$ 
estimates obtained with both the spectroscopic (\S3.1) and photometric 
(\S3.2) methods. The average differences are: $\langle T_{\rm eff}(B - 
V) - T_{\rm eff}$(spec)$\rangle = -65\pm74$\,K, $\langle T_{\rm eff}(V - 
K) - T_{\rm eff}$(spec)$\rangle = -68\pm81$\,K; $\langle\log g(B - V) - 
\log g \rm (spec)\rangle = -0.10\pm0.13$, $\langle\log g(V - K) - \log g 
\rm (spec)\rangle = -0.11\pm0.13$.  The differences observed between the 
two methods are generally consistent with the uncertainties associated 
with the techniques.  The estimation of uncertainties on T$_{\rm eff}$(B 
- V) is $\sim$ 100 K, according to \citet{am99}.  The errors of T$_{\rm 
eff}$(V - K) mainly come from the uncertainties on the $K$ indices, 
which induce a mean error of 90 K, slightly larger than the estimation 
given by \citet{am99}.  The errors of $\log g$ come from the 
uncertainties on parallaxes and mass estimation. The overall estimation 
of errors of $\log g$ is about 0.15 dex, which is consistent with the 
uncertainties estimated with the spectroscopic method.  We also plot 
$\log g$ versus $T_{\rm eff}$ derived with spectroscopic and photometric 
methods in Figure~\ref{fig6}, which shows good consistency between the 
two methods.

\subsection{Infrared Flux Method}


The IRFM is arguably one of the most direct and least model dependent 
techniques to determine effective temperatures in stars (e.g., Blackwell 
\& Shallis 1977; Blackwell et al. 1979, 1980). Our analysis is based on 
the IRFM described in Casagrande et al. (2010, 2014).

The basic idea is to recover for each star its apparent bolometric flux 
and infrared monochromatic flux. One must then compare their ratio to 
that obtained from the same quantities defined on a surface element of 
the star, i.e., the bolometric flux $\sigma T_{\rm eff}^{4}$ and the 
theoretical surface infrared monochromatic flux. For stars hotter than 
$\sim$4200 K (which is the case for our sample) the latter quantity is 
relatively easy to determine because the near infrared region is largely 
dominated by the continuum and depends linearly on $T_{\rm eff}$ 
(Rayleigh-Jeans regime), thus minimizing any dependence on model 
atmospheres. The problem is therefore reduced to a proper derivation of 
stellar fluxes, which can then be rearranged to return the effective 
temperature. Once the apparent bolometric flux and $T_{\rm eff}$ are both 
known, the stellar angular diameter is also trivially obtained.

\begin{figure}
\centering
\includegraphics[width=0.85\textwidth]{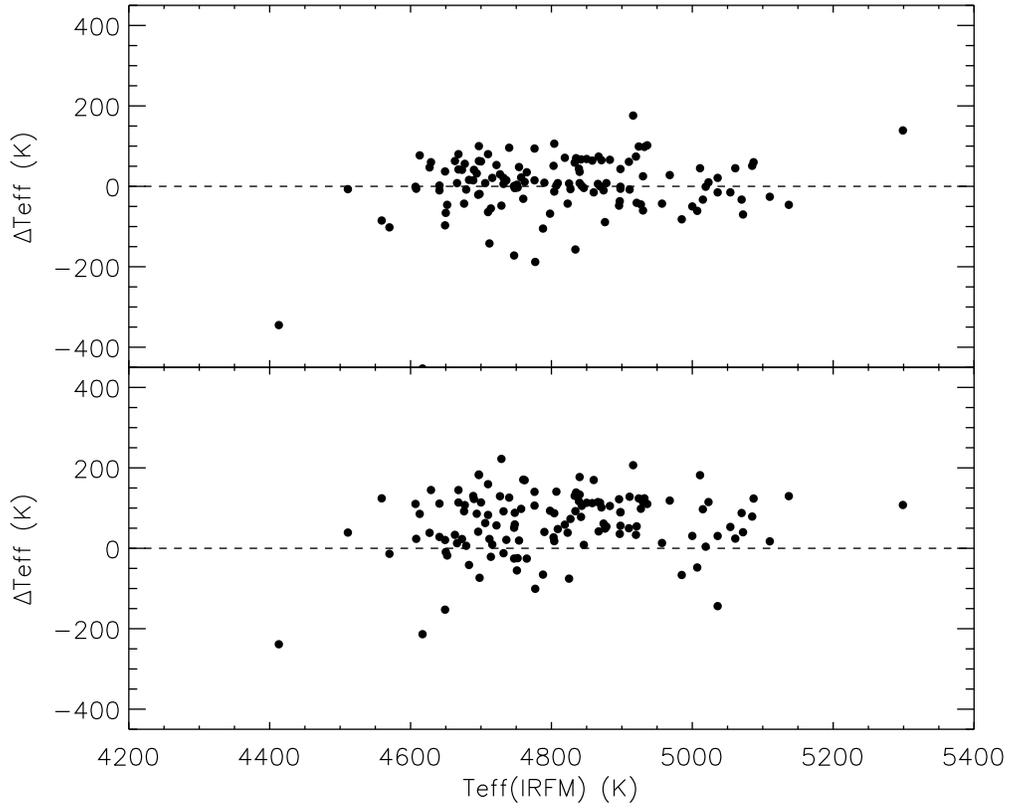}
\caption{Upper panel: T$_{\rm eff}$(IRFM) minus T$_{\rm eff}$(spec) as a 
function of T$_{\rm eff}$(IRFM). Lower panel: T$_{\rm eff}$(IRFM) minus 
T$_{\rm eff}$(B - V) as a function of T$_{\rm eff}$(IRFM).}
\label{fig7}
\end{figure}

In the adopted implementation, the apparent bolometric flux was obtained 
by segments of theoretical model spectrum (for a given $T_{\rm eff}$, 
[Fe/H], and $\log g$) that is normalised by available multi-band 
photometry (i.e. Tycho2 $B_{\rm T}$ $V_{\rm T}$ and 2MASS $JHK_{\rm 
S}$).  The infrared monochromatic flux was derived from 2MASS $JHK_{\rm 
S}$ magnitudes only.  The method critically depends on the availability 
of reliable photometry: some of the brightest stars in 2MASS have 
unreliable magnitudes, and we adopt the same quality cuts as in 
Casagrande et al. (2010) to retain only stars with errors in $J + H + K 
< 0.1$ mag.  These cuts resulted in 34 stars missing an IRFM-derived 
$T_{\rm eff}$ in Table~\ref{phot}.  We used an iterative procedure in 
$T_{\rm eff}$ to cope with the mildly model dependent nature of the 
bolometric correction and surface infrared monochromatic flux.  For each 
star, we used the Castelli \& Kurucz (2004) grid of model fluxes, 
starting with an initial estimate of its effective temperature and 
working at a fixed [Fe/H] and $\log g$ derived from our spectroscopic 
analysis.  The average uncertainty of T$_{\rm eff}$ is about 
80\,K.  We compared the difference of derived T$_{\rm eff}$ with 
spectroscopic, photometric and Infrared Flux method in 
Figure~\ref{fig7}, which shows smaller systematic offset.  The average 
differences are: $<$T$_{\rm eff}$(IRFM) - T$_{\rm eff}$(spec)$>$ = 
1$\pm$150\,K and $<$T$_{\rm eff}$(IRFM) - T$_{\rm eff}$(B - V)$>$ = 
65$\pm$81\,K. 

Uncertainties stemming from the adopted [Fe/H] and $\log g$ were taken 
into account in the error estimate, but their importance is secondary 
since the IRFM has been shown to depend only loosely on those parameters 
(see Casagrande et al. 2006 for a discussion).  This makes the technique 
superior to most spectroscopic methods for determining $T_{\rm eff}$ -- 
provided that reddening is known -- since the effects of $T_{\rm eff}$, 
$\log g$, and [Fe/H] on the latter are usually strongly coupled and the 
model dependence is much more important.  Reddening values described in 
the previous Section were adopted.


\section{Discussion and Conclusions}

\begin{figure}
\centering
\includegraphics[width=0.85\textwidth]{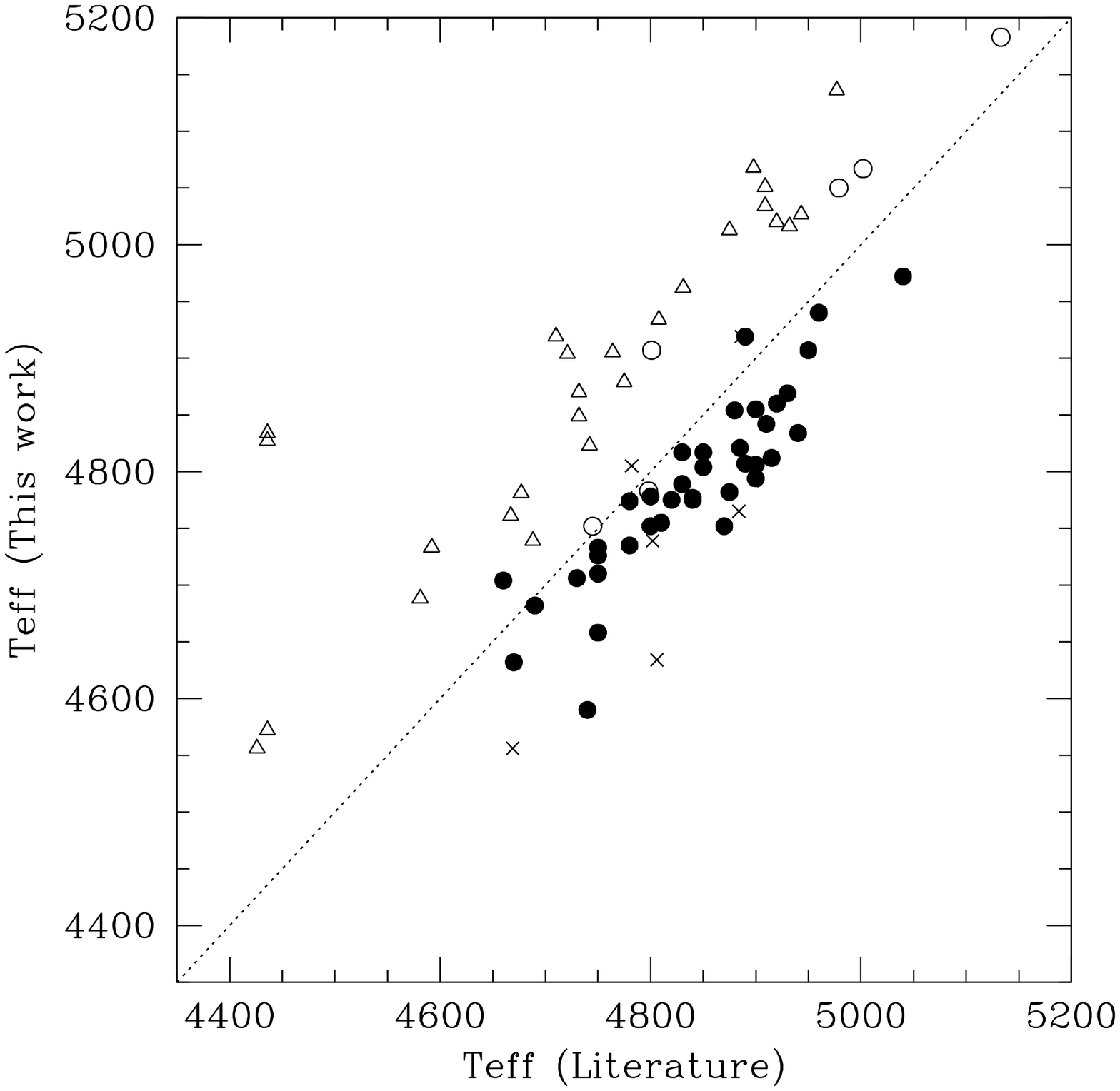}
\caption{Spectroscopic T$_{\rm eff}$(This work) minus T$_{\rm 
eff}$(Literature) for 76 overlapping targets: filled circles -- 
\citet{jones11}, triangles -- \citet{mass08}, crosses -- \citet{luck07}, 
open circles -- \citet{maldonado13}.}
\label{teffcompare}
\end{figure}

\begin{figure}
\centering
\includegraphics[width=0.85\textwidth]{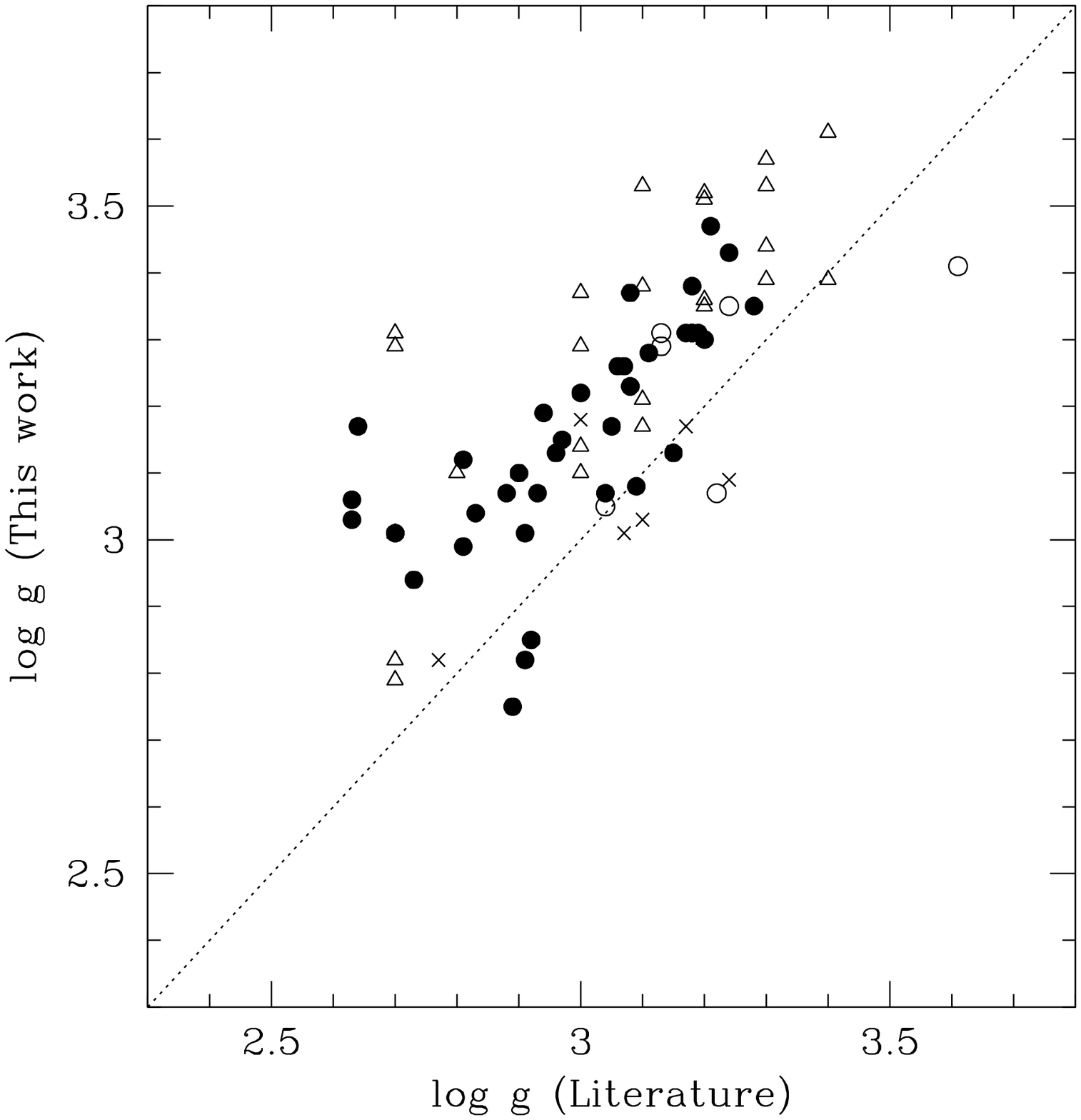}
\caption{Spectroscopic log $g$ (This work) minus log $g$ (Literature) for 
76 overlapping targets: filled circles -- \citet{jones11}, triangles -- 
\citet{mass08}, crosses -- \citet{luck07}, open circles -- 
\citet{maldonado13}.}
\label{loggcompare}
\end{figure}

\begin{figure}
\centering
\includegraphics[width=0.85\textwidth]{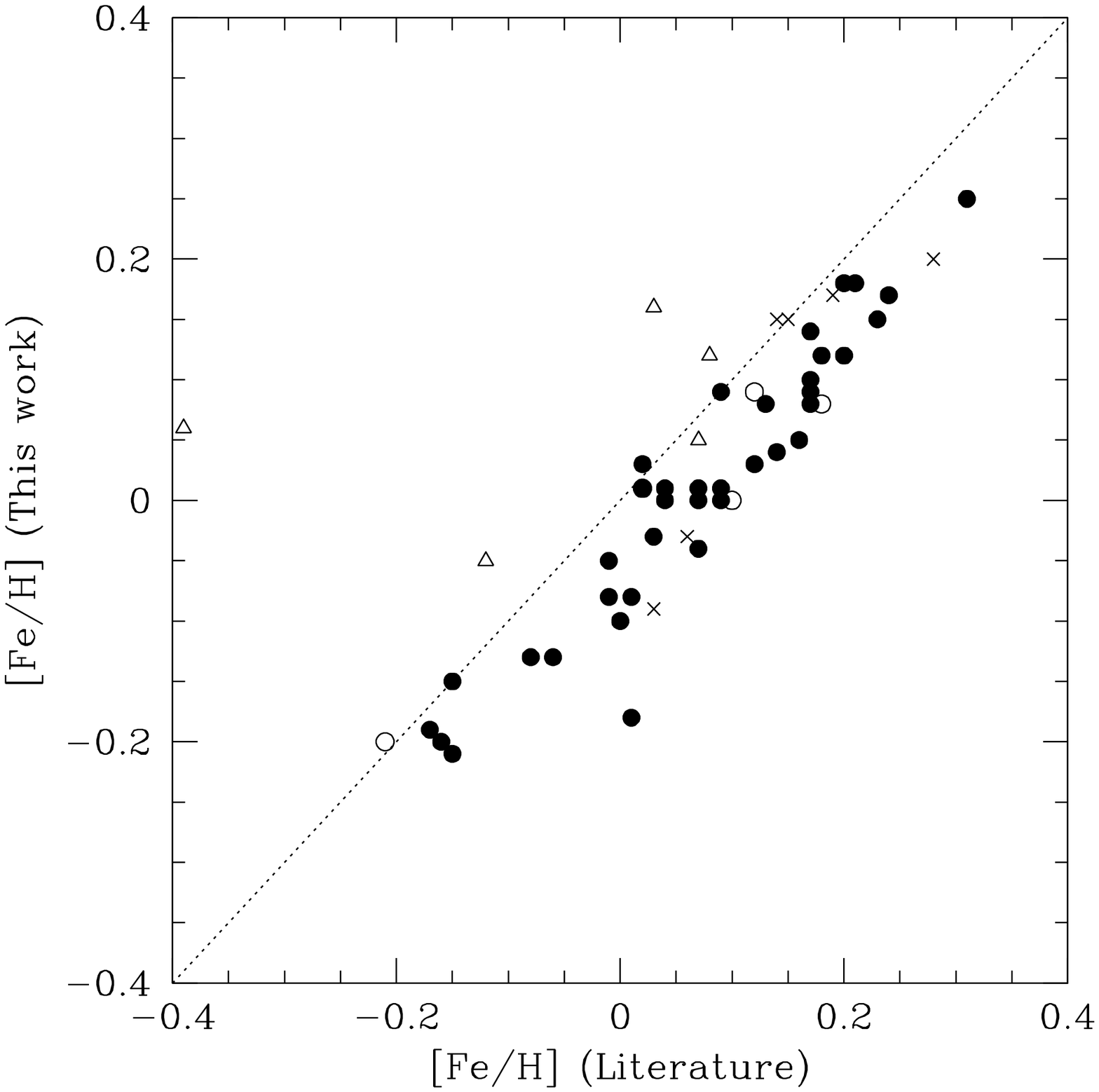}
\caption{Spectroscopic [Fe/H] (This work) minus [Fe/H] (Literature) for 55 
overlapping targets: filled circles -- \citet{jones11}, triangles -- 
\citet{mass08}, crosses -- \citet{luck07}, open circles -- 
\citet{maldonado13}.}
\label{fehcompare}
\end{figure}

Although the PPPS targets are relatively bright stars, less than half of 
them have had fundamental parameter estimates published.  
Table~\ref{compare} gives the previously published spectroscopic 
parameters ($T_{\rm eff}$, $\log g$, and [Fe/H]) for 76 stars from our 
sample.  Our targets have the most overlap, and best agreement with, the 
Southern exoplanet survey of \citet{jones11}.  For $T_{\rm eff}$, we 
have 38 stars in common, with a mean difference of $-52\pm39$\,K.  Good 
agreement is also found for the 6 overlapping stars from \citet{luck07} 
($\Delta\,T=-69\pm$82\,K) and the 6 in common with \citet{maldonado13} 
($\Delta\,T=47\pm$44\,K).  Larger differences are seen for the 26 stars 
in common with \citet{mass08} ($\Delta\,T=146\pm$81\,K).  We 
attribute this difference to the fact that \citet{mass08} computed their 
parameters from published colour indices and metallicities, adopting 
[Fe/H]$=-0.15$ where no published values were available.  That is, 
\citet{mass08} did not derive parameters directly from spectra as this 
work and the others to which we have made comparison.  Results for the 
other spectroscopic parameter comparisons are given in 
Table~\ref{speccomp} and are plotted in 
Figures~\ref{teffcompare}--\ref{fehcompare}.  The overall grand mean 
differences in the parameters are as follows: $\Delta\,T_{\rm eff} = 
22$K, $\Delta\,\log g$ = 0.16 dex, and $\Delta\,$[Fe/H] = $-0.04$\,dex.

We have presented [Fe/H] determinations for 164 evolved stars, 
many of which represent the first such measurements.  As noted in the 
Introduction, the nature of the planet-metallicity correlation (if any) 
remains an unresolved question.  The next logical step is an 
investigation of such a relation for the PPPS sample.  However, a 
complete analysis of the occurrence rate of planets in the PPPS sample 
is beyond the scope of this work, and indeed is premature as we are 
continuing follow-up radial velocity observations for some candidates.  
For example, CHIRON and FEROS data have recently been used 
\citep{jones16} to confirm candidates common between the PPPS and the 
EXPRESS survey of \citet{jones11}.  If we consider the 10 planet hosts 
in this sample (9 published hosts and one in preparation), a K-S test 
comparing the metallicities of the host stars and the 154 non-hosts 
yields $P=0.607$, i.e. a 60.7\% probability that the hosts and non-hosts 
exhibit the same underlying metallicity distribution.  This first-order 
analysis suggests no relation between the star's metallicity and the 
presence of planets, though we caution that no attempt has been made to 
correct for incompleteness, and several promising candidates have not 
been included.  The result of \citet{reffert15}, which did show a 
positive planet-metallicity correlation for evolved stars, remains 
strong evidence due to their careful imposition of uniform planet 
detectability.  We expect to present a similar analysis in a forthcoming 
paper in collaboration with the EXPRESS survey \citep{jones11}.


\acknowledgements

We gratefully acknowledge the UK and Australian government support of 
the Anglo-Australian Telescope through their PPARC, STFC and DIISR 
funding, and travel support from the Australian Astronomical 
Observatory.  This research has made use of NASA's Astrophysics Data 
System (ADS), and the SIMBAD database, operated at CDS, Strasbourg, 
France.  We gratefully acknowledge the efforts of PPPS guest observers 
Hugh Jones, Brad Carter, and Simon O'Toole.  This research has also made 
use of the Exoplanet Orbit Database and the Exoplanet Data Explorer at 
exoplanets.org \citep{wright11}.



\clearpage

\clearpage

\clearpage

\clearpage

\clearpage

\clearpage

\clearpage

\clearpage



\begin{deluxetable}{lrrrr}
\tabletypesize{\scriptsize}
\tablecolumns{5}
\tablewidth{0pt}
\tablecaption{Spectroscopic stellar parameters.}
\tablehead{
\colhead{Star} & \colhead{$T_{\rm eff}$} & \colhead{$\log g$} & 
\colhead{[Fe/H]} & \colhead{$\xi_t$ (\kms)}}
\startdata
\label{spec}
HD~745     &  5160  &  3.29  &  $-$0.04  &  1.19 \\
HD~749     &  4774  &  2.67  &  $-$0.36  &  1.39 \\
HD~1817    &  4630  &  3.02  &     0.03  &  1.16 \\
HD~2816    &  5051  &  3.61  &     0.01  &  1.06 \\
HD~4145    &  4733  &  3.10  &     0.10  &  1.20 \\
HD~4732    &  5008  &  3.32  &  $-$0.04  &  1.21 \\
HD~5676    &  4580  &  2.76  &  $-$0.19  &  1.26 \\
HD~5873    &  4904  &  3.29  &     0.06  &  1.10 \\
\enddata
\end{deluxetable}


\begin{deluxetable}{lrrrrr}
\tabletypesize{\scriptsize}
\tablecolumns{7}
\tablewidth{0pt}
\tablecaption{Stellar parameters from photometric method and IRFM.}
\tablehead{
\colhead{Star} & \colhead{$T_{\rm eff}$$^a$} & \colhead{$\log g$$^a$} & 
\colhead{$T_{\rm eff}$$^b$} & \colhead{$\log g$$^b$} & \colhead{$T_{\rm 
eff}$$^c$} }
\startdata
\label{phot}
HD~745 & 5191 & 3.21 & 5275 & 3.24 & 5299 \\
HD~749 & 4550 & 2.62 & 4645 & 2.75 & 4710 \\
HD~1817 & 4627 & 2.79 & 4588 & 2.76 & 4710 \\
HD~2816 & 5005 & 3.45 & 4988 & 3.43 & 5036 \\
HD~4145 & 4633 & 2.90 & 4782 & 3.06 & --- \\
HD~4732 & 4940 & 3.20 & 4914 & 3.18 & --- \\
HD~5676 & 4588 & 2.75 & 4534 & 2.72 & 4627 \\
HD~5873 & 4808 & 3.17 & 4849 & 3.23 & 4898 \\
\enddata
\tablenotetext{a}{Stellar parameters derived with (B - V).}
\tablenotetext{b}{Stellar parameters derived with (V - K).}
\tablenotetext{c}{Stellar parameters derived from IRFM.}
\end{deluxetable}


\begin{deluxetable}{lrrrrrrr}
\tabletypesize{\scriptsize}
\tablecolumns{8}
\tablewidth{0pt}
\tablecaption{Adopted reddening values for our sample}
\tablehead{
\colhead{Star} & \colhead{$E(B-V)$} & \colhead{$E(V-K)$} & 
\colhead{$A_V$} & \colhead{$BC (B-V)$} & \colhead{$BC (V-K)$} & 
\colhead{$M_{\rm bol} (B-V)$} & \colhead{$M_{\rm bol} (V-K)$} }
\startdata
\label{reddening}
HD~745      &    0.0827080  & 0.2438231  & 0.2563947  &   $-$0.2041 &$-$0.1829  &1.288  &1.309 \\
HD~749      &    0.0130062  & 0.0383422  & 0.0403192  &   $-$0.4597 &$-$0.4045  &1.113  &1.168 \\
HD~1817     &    0.0059238  & 0.0174635  & 0.0183639  &   $-$0.4149 &$-$0.4369  &1.452  &1.430 \\
HD~2816     &    0.0285731  & 0.0842335  & 0.0885766  &   $-$0.2588 &$-$0.2645  &2.457  &2.452 \\
HD~4145     &    0.0095835  & 0.0282522  & 0.0297089  &   $-$0.4112 &$-$0.3426  &1.823  &1.891 \\
HD~4732     &    0.0076328  & 0.0225015  & 0.0236617  &   $-$0.2809 &$-$0.2904  &1.782  &1.772 \\
HD~5676     &    0.0145243  & 0.0428176  & 0.0450253  &   $-$0.4370 &$-$0.4698  &1.515  &1.483 \\
HD~5873     &    0.0221029  & 0.0651595  & 0.0685191  &   $-$0.3317 &$-$0.3149  &2.170  &2.187 \\
\enddata
\end{deluxetable}


\begin{deluxetable}{lrrrr}
\tabletypesize{\scriptsize}
\tablecolumns{5}
\tablewidth{0pt}
\tablecaption{Derived stellar physical parameters}
\tablehead{
\colhead{Star} & \colhead{Mass (\Msun)} & \colhead{$\log L$ (\Lsun)} &
\colhead{Radius (\Rsun)} & \colhead{Age (Gyr)} }
\startdata
\label{masses}
HD~745     &    2.19  &   1.37   &       5.86  & 0.94 \\
HD~749     &    1.32  &   1.49   &       8.06  & 4.72 \\
HD~1817    &    1.12  &   1.37   &       7.33  &  8.09 \\
HD~2816    &    1.48  &   0.94   &       3.86  &  3.02 \\
HD~4145    &    1.26  &   1.17   &       5.46  & 5.55 \\
HD~4732    &    1.65  &   1.22   &       5.45  &  2.18 \\
HD~5676    &    1.02  &   1.43   &       7.30  & 11.08 \\
HD~5873    &    1.31  &   1.05   &       4.62  &  4.69 \\
\enddata
\end{deluxetable}


\begin{deluxetable}{lcccc}
\tabletypesize{\scriptsize}
\tablecolumns{5}
\tablewidth{0pt}
\tablecaption{Spectroscopic parameters of 77 PPPS stars from the literature.}
\tablehead{
\colhead{Star} & \colhead{[Fe/H]} & \colhead{$\log g$} & 
\colhead{$T_{\rm eff}$} }
\label{compare}
\startdata
HD~2816   & \nodata & 3.4     & 4909 & \citet{mass08} \\
HD~4145   &    0.17 & 2.90    & 4750 & \citet{jones11} \\   
          & \nodata & 3.0     & 4592 & \citet{mass08} \\
HD~5873   & $-$0.39 & 3.0     & 4721 & \citet{mass08} \\
HD~5877   &    0.01 & 2.91    & 4750 & \citet{jones11} \\   
HD~6037   &    0.14 & 2.77    & 4669 & \citet{luck07} \\
          & \nodata & 2.7     & 4426 & \citet{mass08} \\
HD~7931   &    0.04 & 3.11    & 4850 & \citet{jones11} \\
\enddata
\end{deluxetable}


\begin{deluxetable}{lllll}
\tabletypesize{\scriptsize}
\tablecolumns{5}
\tablewidth{0pt}
\tablecaption{Mean parameter differences from literature results}
\tablehead{
\colhead{Reference} & \colhead{$N_{stars}$} & \colhead{$T_{\rm eff}$ (K)} 
& \colhead{$\log g$ (cgs)} & \colhead{[Fe/H]} }
\startdata
\label{speccomp}
\citet{jones11}     & 38 & $-52\pm39$\tablenotemark{a}  & $0.17\pm0.13$ & $-0.06\pm0.04$ \\  
\citet{mass08}      & 26 & 146$\pm81$  & $0.23\pm0.15$  & $+0.13\pm0.18$\tablenotemark{b} \\
\citet{luck07}      &  6 & $-69\pm82$  & $-0.01\pm0.11$ & $-0.06\pm0.05$\tablenotemark{c} \\
\citet{maldonado13} &  6 & $47\pm44$   & $0.02\pm0.16$  & $-0.04\pm0.05$\tablenotemark{c} \\
\enddata
\tablenotetext{a}{Computed throughout as (This work) - (Literature value).}
\tablenotetext{b}{$N=5$}
\tablenotetext{c}{$N=6$}
\end{deluxetable}


\clearpage

\end{document}